\documentclass{article}
\usepackage{spconf,amsmath,graphicx}

\usepackage[hidelinks]{hyperref}

\title{MViR: Multi-View Visual-Semantic Representation for Fake News Detection}
\name{Haochen Liang$^{\dag}$, Xinqi Su$^{\dag}$, Jun Wang$^{*}$, Chaomeng Chen, and Zitong Yu$^{*}$\thanks{$^{\dag}$ Equal contribution. $^{*}$ Corresponding authors. \protect\\ \protect\hspace*{1.5em} This work was supported by National Natural Science Foundation of China (Grant No. 62576076), and sponsored by CCF-Tencent Rhino-Bird Open Research Fund.}}
\address{School of Computing and Information Technology, Great Bay University \\
lianghaochenbit@gmail.com, suxinqi@tju.edu.cn, wangjunsdnu@gmail.com \\
gianluigi-chen@bupt.edu.cn, zitong.yu@ieee.org}
\usepackage[T1]{fontenc}
\usepackage[utf8]{inputenc}
\usepackage{microtype}
\usepackage{inconsolata}
\usepackage{graphicx}
\usepackage{times}
\usepackage{latexsym}
\usepackage{fix-cm}
\usepackage{amsmath}
\usepackage{algorithmic}
\usepackage{textcomp}
\usepackage{xcolor}
\usepackage{booktabs}
\usepackage{multirow} 
\usepackage{placeins}
\usepackage{amssymb}
\usepackage{float}
\begin{document}
%
\maketitle
\begin{abstract}
With the rise of online social networks, detecting fake news accurately is essential for a healthy online environment. While existing methods have advanced multimodal fake news detection, they often neglect the multi-view visual-semantic aspects of news, such as different text perspectives of the same image. To address this, we propose a Multi-View Visual-Semantic Representation (MViR) framework. Our approach includes a Multi-View Representation module using pyramid dilated convolution to capture multi-view visual-semantic features, a Multi-View Feature Fusion module to integrate these features with text, and multiple aggregators to extract multi-view semantic cues for detection. Experiments on benchmark datasets demonstrate the superiority of MViR. The source code of FedCoop is available at https://github.com/FlowerinZDF/FakeNews-MVIR.
\end{abstract}
\section{Introduction}
\label{sec:intro}

Fake news refers to deliberately spreading false or misleading information with the aim of deceiving the public, creating confusion, manipulating public opinion, or achieving specific political, economic, or social objectives. Online social networks (OSNs) have increased the convenience of real-time information dissemination, but they also lead to the rapid and widespread dissemination of fake news, causing detrimental effects on the online environment~\cite{zhou2020survey}~\cite{lin2025reliable}. Detecting fake news has thus become a current research hotspot~\cite{su2025dynamic}.

Early works primarily focused on manually extracting features from text content \cite{choudhary2021linguistic}, such as the proportion of negation words, writing style, and language styles. However, traditional methods are inefficient and unable to handle large amounts of data. Therefore, researchers began to focus on deep learning-based automatic fake news detection. Bhattarai et al. \cite{bhattarai2021explainable} captured the lexical and semantic properties of news text. Jin et al. \cite{jin2016novel} detected fake news by leveraging significant disparities in image distributions.

With the development of OSNs \cite{aimeur2023fake}, multimodal fake news which includes text, images, and videos, has emerged. These forms are often more attractive and have a broader reach than traditional unimodal fake news. EANN \cite{wang2018eann} introduces an event discriminator to detect fake news. MVAE \cite{khattar2019mvae} incorporates a multimodal variational autoencoder for multimodal fake news detection. MCAN \cite{wu2021multimodal} designs a co-attention network to better fuse multimodal features.

\begin{figure}[t]
\centering
\includegraphics[width=0.48\textwidth]{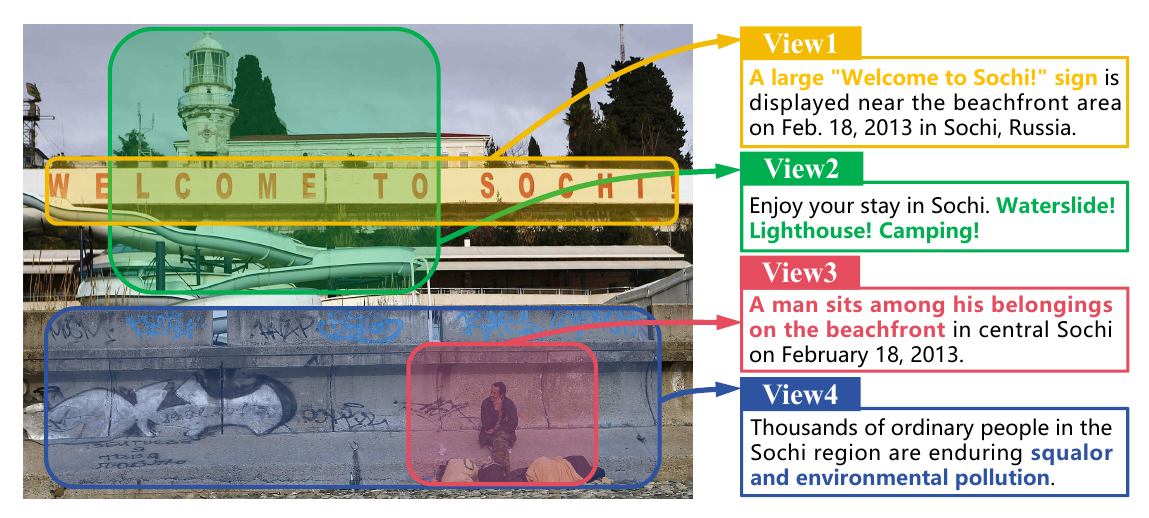}
\vspace{-2.0em}
\caption{Motivation of our proposed MViR. We can see that different news texts describe the same image from various perspectives. For instance, some focus on the background building, others on the sign, and some on the person.} 
\label{fig1}
\vspace{-1.0em}
\end{figure}


\begin{figure*}[t]
\centering
\vspace{-1.0em}
\includegraphics[width=1.0\textwidth]{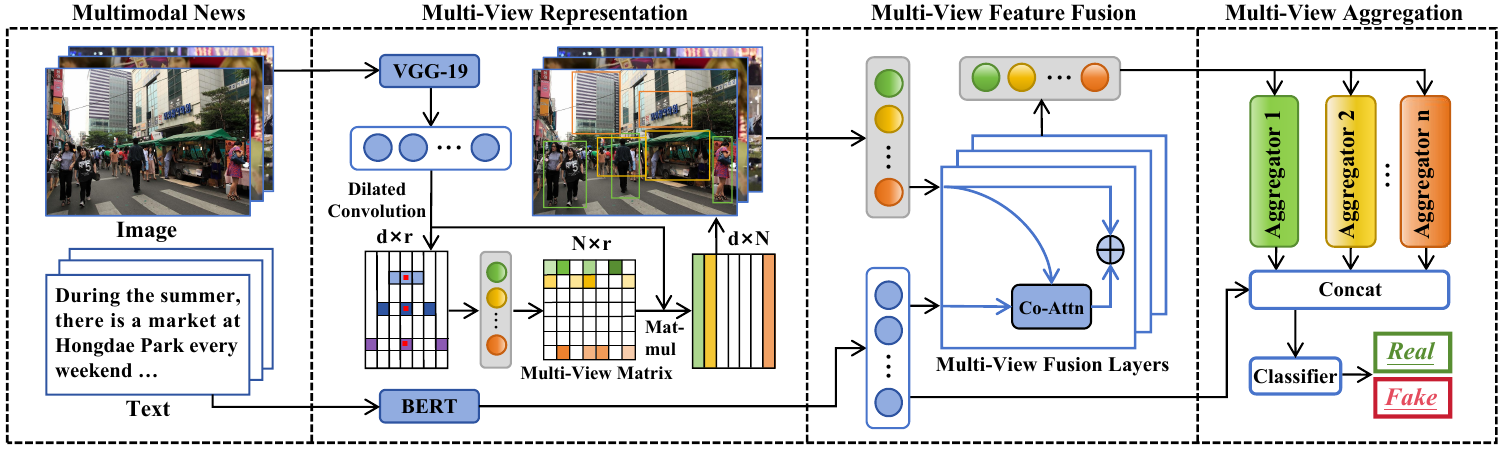}
\vspace{-2.0em}
\caption{The MViR framework consists of three modules: Multi-View Representation (MVR), Multi-View Feature Fusion (MVFF), and Multi-View Aggregation (MVA). It extracts image and text features, learns multi-view visual-semantic representations via MVR, fuses features with MVFF, and uses MVA to generate embeddings and predict fake news probabilities.} 
\label{fig2}
\vspace{-1.5em}
\end{figure*}


However, existing multimodal fake news detection methods struggle to effectively capture the multi-view visual-semantic relationships present in news content. News images often contain information from various perspectives~\cite{ye2025cat+}. For instance, as shown in Figure~\ref{fig1}, the left side illustrates an image associated with fake news, while the right side presents the accompanying text. It is evident that the text reports multiple aspects of the image from different viewpoints, a common phenomenon in real-world news articles. This poses a challenge for previous detection methods \cite{cai2025rehearsal}, as they often fail to consider the multi-view semantics of the content for trustworthy fake news detection.

We propose a multi-view visual-semantic representation for fake news detection (MViR) to address the above issues. Specifically, we propose a multi-view representation module to extract multi-view fine-grained features from images, thereby providing the model with comprehensive multi-view visual-semantic information. Afterward, we use a multi-view feature fusion module to fuse image and text information, further enhancing the representation capability of multi-view features. Finally, we use a Multi-View Aggregation module to process the fused features and extract multi-view semantic cues to enhance fake news detection. Our main contributions include:

(1) We design a multi-view representation module, which can explicitly model the multi-view semantics in news images and capture the multi-view features within the images.

(2) We design a multi-view aggregation module to learn multi-view embeddings and extract multi-view semantic cues for enhanced fake news detection.

(3) Experiments conducted on the widely used datasets show that MViR significantly outperforms previous approaches.

\vspace{-1.0em}

\begin{table*}[htbp]
\vspace{-1.0em}
\centering
\footnotesize
\setlength{\tabcolsep}{7pt} 
\caption{Results on two datasets. The best performance is in bold, while underlining highlights the follow-up.}
\begin{tabular}{lcccccccc}
\hline
\multirow{2}{*}{Dataset} & \multirow{2}{*}{Method} & \multirow{2}{*}{Accuracy} & \multicolumn{3}{c}{Fake News} & \multicolumn{3}{c}{Real News} \\
\cmidrule(lr){4-9}
 &  &  & Precision & Recall & F1 score & Precision & Recall & F1 score \\
\hline
\multirow{8}{*}{Weibo} 
& EANN \cite{wang2018eann} & 0.827 & 0.847 & 0.812 & 0.829 & 0.807 & 0.843 & 0.825 \\
& SAFE \cite{zhou2020similarity} & 0.816 & 0.818 & 0.815 & 0.817 & 0.816 & 0.818 & 0.817 \\
& MCAN \cite{wu2021multimodal} & 0.899 & 0.913 & 0.889 & 0.901 & 0.884 & 0.909 & 0.897 \\
& CAFE \cite{chen2022cross} & 0.840 & 0.855 & 0.830 & 0.842 & 0.825 & 0.851 & 0.837 \\
& FND-CLIP \cite{zhou2023multimodal} & 0.907 & 0.914 & 0.901 & 0.907 & \underline{0.917} & 0.901 & 0.908 \\
& MSACA \cite{wang2024fake} & 0.903 & \underline{0.935} & 0.873 & 0.903 & 0.872 & \underline{0.935} & 0.902 \\
& EVENT-RADAR \cite{ma2024event} & \underline{0.919} & 0.924 & \underline{0.905} & \underline{0.914} & \textbf{0.932} & 0.915 & \underline{0.924} \\
& \textbf{MViR (Ours)} & \textbf{0.924} & \textbf{0.944} & \textbf{0.906} & \textbf{0.920} & 0.906 & \textbf{0.941} & \textbf{0.928} \\
\hline
\multirow{8}{*}{GossipCop} 
& EANN \cite{wang2018eann}  & 0.864 & 0.702 & 0.518 & 0.594 & 0.887 & 0.956 & 0.920 \\
& SAFE \cite{zhou2020similarity} & 0.838 & 0.758 & 0.558 & 0.643 & 0.857 & 0.937 & 0.895 \\
& SPOTFAKE \cite{singhal2020SPOTFAKE+} & 0.858 & 0.732 & 0.372 & 0.494 & 0.866 & 0.962 & 0.914 \\
& CAFE \cite{chen2022cross} & 0.867 & 0.732 & 0.409 & 0.587 & 0.887 & 0.957 & 0.921 \\
& FND-CLIP \cite{zhou2023multimodal} & 0.880 & 0.761 & 0.549 & 0.638 & 0.899 & 0.959 & 0.928 \\
& MSACA \cite{wang2024fake} & \underline{0.887} & \textbf{0.816} & 0.538 & \underline{0.648} & 0.897 & \textbf{0.971} & \underline{0.933} \\
& RaCMC \cite{yu2024racmc} & 0.879 & 0.745 & \underline{0.563} & 0.641 & \underline{0.902} & 0.954 & 0.927 \\

& \textbf{MViR (Ours)} & \textbf{0.895} & \underline{0.784} & \textbf{0.619} & \textbf{0.692} & \textbf{0.914} & \underline{0.963} & \textbf{0.937} \\
\hline
\end{tabular}
\label{tab1}
\vspace{-1.0em}
\end{table*}

\section{Methodology}
\label{sec:format}

As shown in Figure~\ref{fig2}, MViR consists of three main parts: Multi-View Representation (MVR), Multi-View Feature Fusion (MVFF), and Multi-View Aggregation (MVA).


\subsection{Feature  Extraction}

For an image $I$ from multimodal news, we utilize VGG-19~\cite{simonyan2014very} to extract visual features. These features are subsequently projected into a $d$-dimensional space via a fully connected (FC) layer, yielding image features represented as $V = [v_1, v_2, \ldots, v_r] \in \mathbb{R}^{r \times d}$, where \( r \) denotes the number of extracted regions. Similarly, for a text $T$ containing $m$ words, word embeddings are extracted using a pre-trained BERT~\cite{devlin2018bert}, which are mapped to the $d$-dimensional space using a FC layer. Text features are expressed as $T = [t_1, t_2, \ldots, t_m] \in \mathbb{R}^{m \times d}$.

\subsection{Multi-View Representation}

We propose a Multi-View Representation (MVR) module to capture the multi-view semantics of images. Understanding an image from different perspectives means that each region requires different attention from different perspectives. To achieve this, we employ a pyramid dilated convolution~\cite{qu2020context} layer with $K$ parallel kernels to aggregate multi-scale contextual information from the image, which serves as the basis for calculating view-specific importance scores.
\begin{equation}\small
s_i^k = Convd(V, w^k, d^k), \quad k = 1, 2, \dots, K,
\end{equation}
where $w^k$ and $d^k$ denote its kernel size and dilation rate, 
$s_i^k$ denotes the output of the $k$-th kernel. We then concatenate these outputs:
\begin{equation}\small
    s_i = Concat(s_i^1, ..., s_i^K),
\end{equation}
where $Concat(\cdot)$ denotes the concatenation of vectors. Afterward, a FC followed by a softmax activation is applied to compute the multi-view matrix $\hat{S} = [\hat{s}_1; ... ; \hat{s}_r] \in \mathbb{R}^{r \times N}$, where $N$ is the number of views, and $\hat{s}_i$ is the $i$-th row vector. The above process can be summarized as follows:
\begin{equation}\small
    \hat{s}_{ij} = \frac{\exp\left((W_s s_i + b_s)_j\right)}{\sum_{j=1}^r \exp\left((W_s s_i + b_s)_j\right)},
\end{equation}
where $(W_s s_i + b_s) \in \mathbb{R}^{r \times N} $, \( \hat{s}_i \in \mathbb{R}^N \) represents the importance scores of the \( i \)-th region over \( N \) views, \( W_s \in \mathbb{R}^{N \times d} \) and \( b_s \in \mathbb{R}^{1 \times N} \) are the learnable weights and bias, respectively. Finally, the image features can be summarized into a multi-view representation \( \mathbf{V}^* \in \mathbb{R}^{N \times d} \) as follows:
\begin{equation}\small
\mathbf{V}^* = \mathbf{\hat{S}}^T \mathbf{V}.
\end{equation}

\subsection{Multi-View Feature Fusion}

We propose a Multi-View Feature Fusion (MVFF) module that combines multi-view image features with the corresponding text features, further enhancing their representational capacity. MVFF consists of $l$ multi-view fusion layers, each containing a co-attention  mechanism~\cite{lu2019vilbert} and a feed-forward network (FFN). Both components are enclosed by a residual connection and followed by layer normalization. The co-attention extends the standard multi-head attention by using queries ($Q$) from one modality and keys ($K$) and values ($V$) from another. In our approach, $Q$ comes from the multi-view features of the image or fused features, while $K$ and $V$ come from the text.
\begin{equation}\small
    Q_i = \mathbf{V}^*W_i^Q, \quad K_i = TW_i^K, \quad V_i = TW_i^V ,
\end{equation}
where $W_i^Q, W_i^K, W_i^V \in \mathbb{R}^{1 \times d_h}$ are the projection matrices for the $i$-th head, $H$ denotes the number of heads, $d_h = d/H$ is the dimension of the output feature of each head. The calculation of the co-attention can be presented as follows:
\begin{equation}\small
    MultiHead(Q, K, V) = Concat(h_1,h_2,...,h_H)W^O + V,
\end{equation}
where $W^O$ is learnable weights, and $h_i = Att(Q_i,K_i, V_i)$. $Att$ denotes scaled-dot product attention, defined as follows:
\begin{equation}\small
    Att(Q_i,K_i, V_i) = softmax\left(\frac{Q_i K_i^\top}{\sqrt{d_h}}\right)V_i.
\end{equation}
To enhance the representational capacity of the fused features, the output of the co-attention is processed through an FFN. It is implemented as a two-layer multi-layer perceptron (MLP) with a ReLU activation function applied between the layers. The above process can be summarized as follows:
\begin{equation}\small
   \mathbf{X} = FFN(MultiHead(Q, K, V)) \oplus Q  ,
\end{equation}
where $\mathbf{X} \in \mathbb{R}^{N \times d} $ denotes the features after fusion, $\oplus$ represents the fusion operation, e.g., vector concatenation or elementwise add.

\subsection{Multi-View Aggregation}

The uniqueness of the Multi-View Aggregation (MVA) module lies in its use of multiple aggregators to generate a set of embeddings from the fused features, explicitly modeling multi-view features. This allows for the evaluation of news authenticity through multi-view semantic cues. Specifically, after obtaining a fused feature set $\{\mathbf{x}_n\}_{n=1}^N$, a series of feature aggregators $\{f_n^v\}_{n=1}^N$ are used to aggregate $\{\mathbf{x}_n\}_{n=1}^N$ into a set of semantic cues embeddings $\{\hat{\mathbf{x}}_n\}_{n=1}^N$:
\begin{equation}\small
    \hat{\mathbf{x}}_n = f_n^v\left(\{\mathbf{x}_n\}_{n=1}^N\right),
\end{equation}
where $\hat{\mathbf{x}}_n \in \mathbb{R}^d$. Each \( \hat{\mathbf{x}}_n \) represents the semantic cues of a set of views. $N$ is the number of views. 

Next, we combine the semantic cues from each view with the text features and use a decision network to assess whether the news is true or fake. We employ only a single aggregator to obtain the text embedding:
\vspace{-0.5em}
\begin{equation}\small
    z_n = \max \left( 0, W_f \text{Concat}(\hat{\mathbf{x}}_n, f^t(T)) \right), 
\end{equation}
\begin{equation}\small
    \hat{y} = \max_{n=1}^{N} \left( \text{softmax} \left( z_n W_l \right) \right),
\end{equation}
where \(\hat{y}\) represents the probability of the news being fake, $W_f$ represents the parameters of the fully connected layers, and $W_l$ represents the parameters of the linear layer within the softmax function. For multiple semantic cues from different views of news, if any of these cues is detected as fake, the news is classified as fake. 

\subsection{Objective Function}

We leverage cross-entropy to measure the classification loss and train our model:
\begin{equation}\small
\mathcal{L}=\sum_{\mathrm{i}=1}^\mathcal{N}-\left[y_{\mathrm{i}} * \log \left(\hat{y}_{\mathrm{i}}\right)+\left(1-y_{\mathrm{i}}\right) * \log \left(1-\hat{y}_{\mathrm{i}}\right)\right],
\end{equation}
where $\mathcal{N}$ denotes the number of news reports, $y_{\mathrm{i}}$ represents the ground-truth label of the $i$-th news. Labels 0 and 1 refer to real news and fake news, respectively.

\section{Experiments}
\label{sec:pagestyle}

\subsection{Datasets and Experimental Settings}

Our model, implemented in PyTorch 2.3.1, was trained on a single NVIDIA Tesla A100 GPU. For text features, we utilized bert-base-chinese (max length: 160) for the Weibo dataset and bert-base-uncased (max length: 394) for GossipCop. Images were resized to $224 \times 224$ pixels for VGG-19 feature extraction. The feature dimension $d$ for both modalities was set to 256, with 4 attention heads and a 0.5 dropout rate. Training employed the AdaBelief optimizer for 50 epochs, using a batch size of 32 and an initial learning rate of 1e-4.

Experiments were conducted on two real-world datasets: Weibo \cite{jin2017multimodal} (9,528 news items) and GossipCop \cite{shu2020fakenewsnet} (12,840 news items). For fair comparison, we adhered to the data-splitting and processing protocols of prior works~\cite{wu2021multimodal,ying2023bootstrapping}.

A core component is the Pyramid Dilated Convolutional Layers, with configurations detailed in Table~\ref{tab2}. Here, $k$ denotes the kernel index; $w^k$, $d^k$, and $s^k$ represent the kernel size, dilation rate, and output channels, respectively. This design effectively expands the receptive field with increasing dilation rates without compromising spatial resolution.

\begin{table}[t]
\vspace{-1.0em}
\centering
\footnotesize
\setlength{\tabcolsep}{5pt}
\vspace{-0.5em}
\caption{Configurations of pyramid dilated convolution.}
\begin{tabular}{cccccccc}
\hline
$k$ & 1 & 2 & 3 & 4 & 5 & 6 & 7 \\ \hline
$w^k$ & 1 & 3 & 3 & 3 & 5 & 5 & 5 \\
$d^k$ & 1 & 1 & 2 & 3 & 1 & 2 & 3 \\
$s^k$ & 256 & 128 & 128 & 128 & 128 & 128 & 128 \\ \hline
\end{tabular}
\vspace{-0.5em}
\label{tab2}
\end{table}

\begin{table}[t]
\centering
\vspace{-0.5em}
\caption{Ablation study on two datasets.}
\resizebox{\linewidth}{!}{ 
\begin{tabular}{@{}cc@{}ccc@{}}
\hline
\multirow{2}{*}{}        & \multirow{2}{*}{Method} & \multirow{2}{*}{Accuracy} & \multicolumn{2}{c}{F1 score}    \\ \cline{4-5} 
                         &                         &                           & Fake News      & Real News      \\ \hline
\multirow{6}{*}{Weibo}   & MViR (Ours)                   & \textbf{0.924}            & \textbf{0.920} & \textbf{0.928} \\ \cline{2-5} 
                         & w/o MVR            & 0.901                     & 0.893          & 0.909          \\ 
                         & w/o MVFF            & 0.894                     & 0.884          & 0.902          \\ 
                         & w/o MVA            & 0.907                     & 0.904          & 0.909          \\ \cline{2-5} 
                         & Max Probability(Real)    & 0.912                     & 0.908          & 0.915          \\
                         & Average Probability    & 0.918                     & 0.915          & 0.921          \\\hline
\multirow{6}{*}{GossipCop} & MViR (Ours)                    & \textbf{0.895}            & \textbf{0.692} & \textbf{0.937} \\ \cline{2-5} 
                         & w/o MVR            & 0.883                     & 0.648          & 0.930          \\
                         & w/o MVFF            & 0.881                     & 0.667          & 0.927          \\ 
                         & w/o MVA            & 0.886                     & 0.639          & 0.932          \\\cline{2-5}
                         & Max Probability(Real)    & 0.874                     
                         & 0.655          & 0.915          \\
                         & Average Probability    & 0.884                     
                         & 0.653          & 0.929          \\ \hline
\end{tabular}
}
\label{tab3}
\vspace{-1.0em}
\end{table}

\subsection{Performance Comparison}

Table~\ref{tab1} shows the performance comparison between MViR and the baseline methods. MViR demonstrated excellent performance across all metrics, including accuracy, precision, recall, and F1 score. MViR achieved an average accuracy of 92.4\% on the Weibo dataset and 89.5\% on the GossipCop dataset, outperforming the best existing models by 1.9 and 1.7 percentage points.

While many methods, such as MCAN, detect fake news by fusing multimodal features, they do not consider the multi-view characterization of fake news. In contrast, MViR effectively captures the multi-view features of images, thus improving the performance of multimodal fake news detection.

\subsection{Ablation Study}\label{as}

To evaluate the contribution of each component in MViR, we removed the MVR, MVFF, and MVA modules individually. Table~\ref{tab3} shows that removing any module leads to a significant performance drop, confirming that the components complement each other to improve fake news detection. Additionally, we compared decision networks using maximum real news probability, average probability, and maximum fake news probability. The results indicate that the network using maximum fake news probability performs best, as it can capture more reliable key features of fake news.

\subsection{Parameter Sensitivity Analysis}

\noindent \textbf{Impact of the Number of MVFF Layers: }As shown in Table~\ref{tab4}. The experimental results demonstrate that appropriately increasing the number of layers can promote interactions between modality features, thereby enhancing model performance. However, when the number of layers becomes larger, further increases result in performance degradation, likely due to excessive model complexity leading to overfitting.

\begin{figure}[t]
\centering
\includegraphics[width=0.47\textwidth]{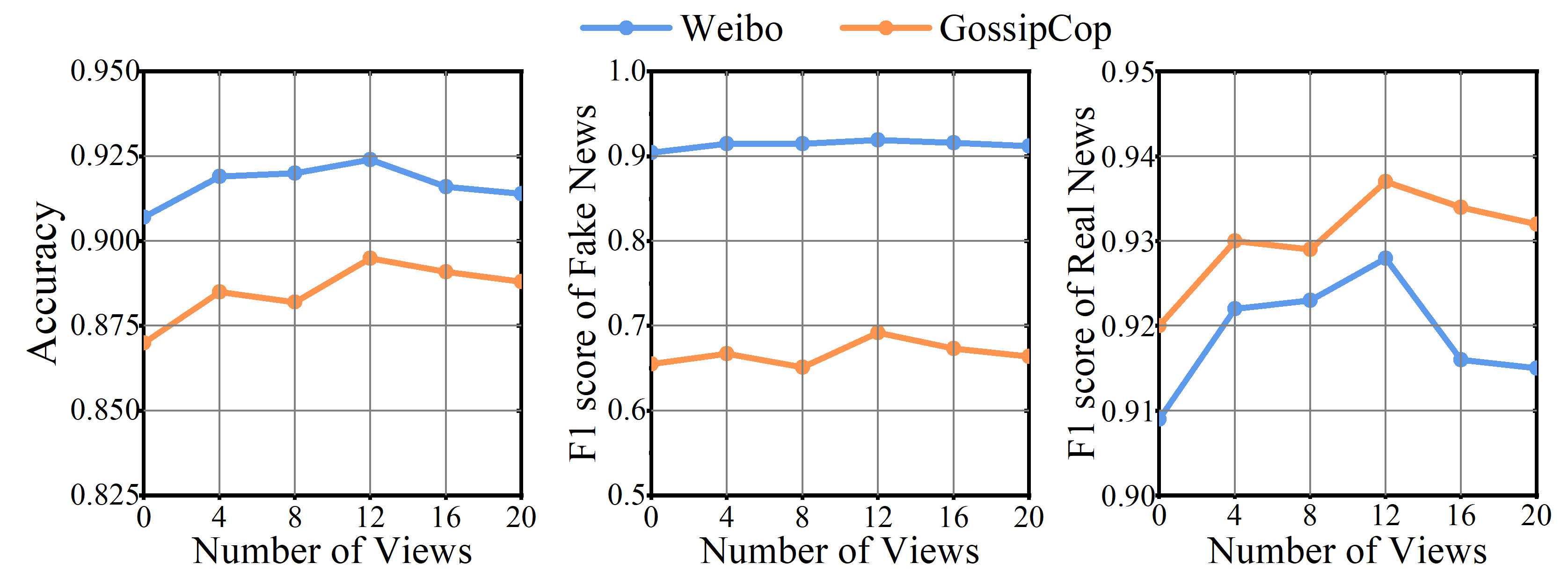}
\vspace{-0.5em}
\caption{Analysis for different numbers of views.} 
\label{fig3}
\vspace{-0.5em}
\end{figure}

\noindent \textbf{Impact of the Number of views: }We also analyzed the impact of the number of viewpoints on model performance, as shown in Figure~\ref{fig3}. It can be observed that appropriately increasing the number of viewpoints helps capture richer details and more diverse features, thereby improving performance. However, when the number of viewpoints increases further, the model's performance may decline due to the introduction of redundant information, which could affect its generalization ability. In our experiments, MViR performed best with 12 viewpoints.

\section{Conclusion}

In this work, we propose MViR, a novel framework for fake news detection using multi-view visual-semantic representations. Our approach includes a multi-view representation module to extract visual-semantic features from images, a feature fusion module to combine image and text features, and a Multi-View Aggregation module to learn multi-view embeddings. Experiments on two benchmark datasets show that MViR outperforms existing state-of-the-art methods.

\begin{table}[t]
\vspace{-1.5em}
\centering
\footnotesize
\caption{Analysis for different numbers of MVFF layer.}
\resizebox{\linewidth}{!}{ 
\begin{tabular}{ccccc}
\hline
\multirow{2}{*}{}        & \multirow{2}{*}{Layers} & \multirow{2}{*}{Accuracy} & \multicolumn{2}{c}{F1 score}    \\ \cline{4-5} 
                         &                         &                           & Fake News      & Real News      \\ \hline
\multirow{3}{*}{Weibo}   & 2            & 0.912                     & 0.908          & 0.915 \\
                         & 3            & \textbf{0.924}            & \textbf{0.920} & \textbf{0.928}          \\ 
                         & 4            & 0.917                     & 0.915          & 0.921          \\ \hline
\multirow{3}{*}{GossipCop} & 2                    & 0.884                     & 0.648          & 0.930 \\
                         & 3            & \textbf{0.895}            & \textbf{0.692} & \textbf{0.937}          \\
                         & 4            & 0.891                     & 0.674          & 0.935          \\ \hline
\end{tabular}
}
\label{tab4}
\end{table}

{
    \vspace{-0.5em}
    \let\oldthebibliography\thebibliography
    \renewcommand\thebibliography[1]{%
      \oldthebibliography{#1}%
      \setlength{\itemsep}{0pt} 
      \setlength{\parskip}{0pt} 
    }
    \vspace{-0.5em}
    \bibliographystyle{IEEEbib}
    \bibliography{main}
}

\end{document}